\documentclass[aps,prd,twocolumn,showkeys,groupedaddress,nofootinbib]{revtex4-1}
\usepackage{graphicx}
\usepackage{subfig}
\usepackage{filecontents}
\begin{document}

\preprint{Proceedings of the $22^{nd}$ International Spin Symposium}

\title{COMPASS measurement of the $P_T$ weighted Sivers asymmetry}



\author{Franco Bradamante}
\affiliation{INFN Sezione di Trieste, Trieste, Italy}

\collaboration{on behalf of the COMPASS Collaboration}
\noaffiliation

\date{\today}

\begin{abstract}
The SIDIS  transverse spin asymmetries weighted with powers of $P_T$, 
the hadron transverse momentum in the $\gamma N$ reference system, 
have been introduced
already twenty years ago and are considered quite interesting. 
While the amplitudes of the
modulations in the azimuthal distribution of the hadrons are the 
ratios of convolutions
over transverse momenta of the transverse--momentum dependent parton
distributions and of the corresponding 
fragmentation functions, and can be evaluated analytically only 
making assumptions on 
the transverse--momentum 
dependence of these functions, the weighted asymmetries allow 
to solve the
convolution integrals over transverse--momenta without those assumptions.
Using the high statistics data collected in 2010 on transversely 
polarized proton target
COMPASS has evaluated in x-bins the $P_T$ weighted Sivers asymmetry 
which is proportional
to the product of the first transverse moment of the Sivers 
function and of the
fragmentation function. The results are compared to the 
standard unweighted
Sivers asymmetry.
\end{abstract}
%\pacs{}
%\keywords{...}



\maketitle

In the generalized parton model, admitting a finite intrinsic transverse
momentum $\vec{k}_T$ for the quarks, a total of eight Transverse Momentum Dependent 
(TMD) Parton Distribution Functions (PDFs) are needed for a full description 
of the structure of the nucleon at leading twist 
\cite{Kotzinian:1994dv,Mulders:1995dh,Bacchetta:2006tn}.
These functions lead to asymmetries in the azimuthal distributions of 
hadrons produced in Semi-Inclusive measurements of Deeply Inelastic 
Scattering (SIDIS) processes off polarised and unpolarised nucleons
and can be disentangled by measuring the amplitudes of the
different angular modulations. 
Among the eight TMD PDFs, the T-odd Sivers function \cite{Sivers:1989cc} 
is of particular 
interest. 
This function arises from a correlation between the transverse 
momentum 
of an unpolarised quark in a transversely polarised nucleon and 
the nucleon spin. It is responsible for the Sivers asymmetry, $A_{Siv}$, 
which is proportional to the convolution of the Sivers function
$f_{1T}^{\perp }$ and of the 
unpolarised fragmentation function $D_1$.
This asymmetry consists in a sin $\Phi_{Siv}$ 
modulation in the number of the produced hadrons, where 
$\Phi_{Siv} = \phi_h-\phi_S$ is the 
difference of the azimuthal angle of the hadron and of the nucleon spin. 
The azimuthal angles are defined in a reference system in which the z axis 
is the virtual photon direction  and the xz plane is the lepton scattering 
plane. 

The Sivers effect was experimentally observed in SIDIS  on 
transversely polarised proton targets, first by the HERMES 
Collaboration \cite{Airapetian:2004tw,Airapetian:2009ae} 
and then by the COMPASS Collaboration 
\cite{Alekseev:2010rw,Adolph:2012sp,Adolph:2014zba}. 
A small effect was also observed at JLab for positive hadrons produced on a 
$^3$He target \cite{Qian:2011py}, while the COMPASS measurements 
on a transversely  polarised deuteron target 
\cite{Alexakhin:2005iw,Ageev:2006da,Alekseev:2008aa} 
gave asymmetries compatible with zero. 
Combined analysis of the proton and deuteron data soon allowed for first 
extractions of the Sivers function for u- and d-quarks 
\cite{Efremov:2003tf,Vogelsang:2005cs,Collins:2005ie,Anselmino:2012aa}, 
which turned out to be different from zero, with similar strength and 
opposite in sign, a most important result in TMD physics. 
In the standard “Amsterdam” notation the Sivers asymmetry can be written 
as 
\begin{eqnarray}
A_{Siv}(x,z) &=&
\frac{\sum_q e_q^2 x f_{1T}^{\perp  \, q}(x) \otimes D_1^q(z)}
     {\sum_q e_q^2 x f_{1}^q(x) \cdot D_1^q(z)}
\label{eq:sa}
\end{eqnarray}
where $x$ is the Bjorken variable,
$z$ is the fraction of the available energy carried  
by the hadron, and
$\otimes$ indicates a convolution over the transverse momenta
of the Sivers function $f_{1T}^{\perp}$ 
and the fragmentation function $D_1$. 
In all those analyses, due to the presence of the convolution, some functional 
form had to be assumed
for the transverse momentum dependence of both the quark distribution 
functions and of the fragmentation
functions.
In most of the analyses these functions were assumed to be Gaussian, and 
as a result the first transverse  moment of the 
Sivers function 
\begin{eqnarray}
f_{1T}^{\perp (1)}(x) = \int d^2  \vec{k}_T 
\, \frac{k_T^2}{2 M^2} \,  
\, f_{1T}^{\perp}(x, k_T^2) 
\label{k_T_moment}
\end{eqnarray}
could be determined from the asymmetry data. 

Already twenty years ago an alternative method was proposed 
to determine $f_{1T}^{\perp (1)}$
without making any assumption on the functional form
neither of the distribution functions nor of the fragmentation functions.
The method consists in measuring asymmetries weighted by the 
transverse momentum of the hadron $P_T$
 \cite{Kotzinian:1995cz,Kotzinian:1997wt,Boer:1997nt} as described in 
the following.
For some reasons the method was not pursued: 
the only results (still preliminary) came from HERMES 
\cite{Gregor:2005qv} and have already 
been used to extract the transverse moment of the Sivers function 
\cite{Collins:2005ie}. 
Recently, much interest has been dedicated again to the weighted asymmetries 
(see, f.i. \cite{Kang:2012ns}).
In this contribution the first COMPASS results  \cite{Sbrizzai2016} 
on the weighted Sivers
asymmetries from the data collected in 2010 on a transversely 
polarized proton target are presented.\\

Keeping only the relevant terms, the SIDIS cross-section can be written as
\begin{eqnarray} 
\frac{d\sigma}{dx dy dz  d\Phi_{Siv}} &= &
\frac{\alpha^2}{s}\,
\frac{1-y+ y^2/2}{x^2 y^2} \nonumber \\
& \times & \left[ F_{UU} + {S}_{T} \,F_{UT}^{\sin \Phi_{Siv}} \, \sin \Phi_{Siv}+... 
\right] \; \; \; \;
\label{eq:sidiscs}
\end{eqnarray} 
where $\alpha$ is the fine structure constant, $s=Q^2/xy$ is the center-of-mass
energy squared, and terms of the order of $\gamma^2 = (2Mx/Q)^2$, where
$M$ is the nucleon mass, have been neglected. 
    The quantities $y$ and $Q^2$ are the fraction 
of  lepton energy carried away by the virtual photon
and the photon virtuality 
respectively.
        
The Sivers asymmetry $A_{Siv}$ is the ratio of the structure functions 
\begin{eqnarray}
&F_{UT}^{sin \Phi_{Siv}} &=\int d^2\vec{P}_T \, P_T F(P_T^{\, 2}) \nonumber \\
&F_{UU}  \; \; \; \;  \; \; &= \int d^2\vec{P}_T 
\int d^2\vec{k}_T \int d^2\vec{p}_{\perp} \delta^2 \, (\vec{P}_T-\vec{p}_{\perp}-z\vec{k}_T) \, \nonumber \\
&&  \; \;  \;  \; \;  \; \times \; f_{1}(k_T^2) D_1(p_{\perp}^2)
\end{eqnarray}
where
 \begin{eqnarray}
F(P_T^{\, 2}) &=&
\int d^2\vec{k}_T \int d^2\vec{p}_{\perp} \, \delta^2 (\vec{P}_T-\vec{p}_{\perp}-z\vec{k}_T) \nonumber \\
&&  \; \;  \;  \; \;  \; \times \; \frac{\vec{P}_T \cdot k_{T}}{M  P_T^{\, 2}} 
f_{1T}^{\perp}(k_T^2) D_1(p_{\perp}^2) 
\end{eqnarray}
 and $\vec{p}_{\perp}$ is the hadron transverse momentum 
relative to the fragmenting quark. 
Although it has not been explicitly  indicated
 the quark distribution functions depend also on $x$ and $Q^2$ and 
 the fragmentation functions on $z$ and $Q^2$.
Also, the sum over the quark flavor is not indicated.

The convolution which appears in $F_{UU}$ can be easily calculated (the usual 
relation among the transverse momenta of the quark and of the hadron is 
guaranteed by the delta function), but in general this is not the case for 
$F_{UT}^{sin \Phi_{Siv}}$. 
It is possible to calculate it by assuming specific functional forms for 
the transverse momentum dependence of 
$f_{1T}^{\perp}$ and $D_1$, and a common choice is the Gaussian “ansatz”
% f_1(x, k_T^2) = f_1(x) \, \frac{e^{- k_T^2/\langle k_T^2 \rangle}}{\pi 
% \langle k_T^2 \rangle}\,, \;
% \label{tmd_gauss1} \\
\begin{eqnarray}
f_{1T}^{\perp}(x, k_T^2) &=& f_{1T}^{\perp}(x) \, \frac{e^{- k_T^2/\langle k_T^2 \rangle_S}}{\pi 
\langle k_T^2 \rangle_S}\,, \; \; \; \;
\nonumber \\
D_1(z, p_{\perp}^2)  &=& D_1(z) \, \frac{e^{- p_{\perp}^2/\langle p_{\perp}^2 \rangle}}{\pi 
\langle p_{\perp}^2 \rangle}\,,  
\label{tmd_gauss3} 
\end{eqnarray}
which gives 
\begin{eqnarray}
A_{Siv,G}(x,z) &=& a_G 
\frac{\sum_q e_q^2 x f_{1T}^{\perp \, (1) \, q}(x) \cdot D_1^q(z)}
     {\sum_q e_q^2 x f_{1}^q(x) \cdot D_1^q(z)} \, .
\label{eq:gsa}
\end{eqnarray}
where
$a_G = \sqrt{\pi}M/\sqrt{\langle k_T^2 \rangle _S + 
\langle p^2_{\perp} \rangle /z^2}$.
From eq. (\ref{eq:gsa}) it is clear that in the Gaussian ansatz
the first moment of the Sivers function $f_{1T}^{\perp \, (1)}$
can be evaluated directly from the measured Sivers asymmetries
using some assumption for the transverse momenta which appear in $a_G$.

As stated above, the first moment of the Sivers function
% $f_{1T}^{\perp(1)}$
can be accessed  without making any 
hypothesis on the specific functional forms for $f_{1T}^{\perp}$ and 
$D_1$ by measuring the $P_T$ weighted Sivers asymmetries. 
These asymmetries are obtained by weighting the events 
by the measured transverse momentum of
the hadron.
Only the spin-dependent
part of the cross-section has to be weighted,
leaving unweighted the unpolarised cross-section.
In this work the weighting is done with $P_T/zM$, where $M$ is the 
nucleon mass.
% , but it is planned also to weight only by $P_T/M$. 
After some algebra one gets the simple result
 \begin{eqnarray}
F_{UT}^{sin \Phi_{Siv}, \, w}=\int d^2\vec{P}_T \frac{P_T^{\, 2}}{zM}\,  F(P_T^{\, 2})
=2\,f_{1T}^{\perp (1)} \, D_1
\end{eqnarray}
namely the convolution becomes the product of the first transverse 
moment of the Sivers function and the fragmentation function $D_1$, 
so that the weighted Sivers asymmetry  is
\begin{eqnarray}
A_{Siv}^{w}(x,z) &=&
2\frac{\sum_q e_q^2 x f_{1T}^{\perp \, (1) \, q}(x) \, D_1^q(z)}
     {\sum_q e_q^2 x f_{1}^q(x) \, D_1^q(z)} \,.
\label{eq:wsa}
\end{eqnarray}
An interesting remark is that, knowing the unpolarised distributions and 
fragmentation functions,
both  $f_{1T}^{\perp \, (1) \, u}$ and  $f_{1T}^{\perp \, (1) \, d}$ could
easily be obtained following the procedure of Ref. \cite{baronespin}.
Also, assuming 
u-dominance for positive hadrons produced
on a proton target, the fragmentation function
cancels out and the asymmetry simply becomes
\begin{eqnarray}
A_{Siv}^{w}(x,z) &\simeq&
2\frac{f_{1T}^{\perp \, (1) \, u}(x)}
     {f_{1}^u(x) } \,.
\label{eq:wsag}
\end{eqnarray}

\par
The data we used for this analysis are the data collected in 2010 with 
the transversely polarised proton target. 
Since the comparison with the “standard” Sivers asymmetry is also important, 
the data production and all the cuts to select the muons and the 
hadrons are the same as for the published data \cite{Adolph:2012sp}. 
In particular, the selected phase space is defined by $0.004 < x < 0.7$, 
$Q^2 > 1$ GeV/c$^2$,  $0.1<y < 0.9$,  $W > 5$ GeV/c$^2$, $P_T > 0.1$ GeV/c, 
and $z > 0.2$.
Presently, the asymmetry has been measured only as a function of $x$ and
the results have been extracted in the nine $x$-bins of ref. \cite{Adolph:2012sp}. 

\begin{figure*}[tb]
% \begin{minipage}{.99\textwidth}
\centering
\includegraphics[width=0.98\textwidth]{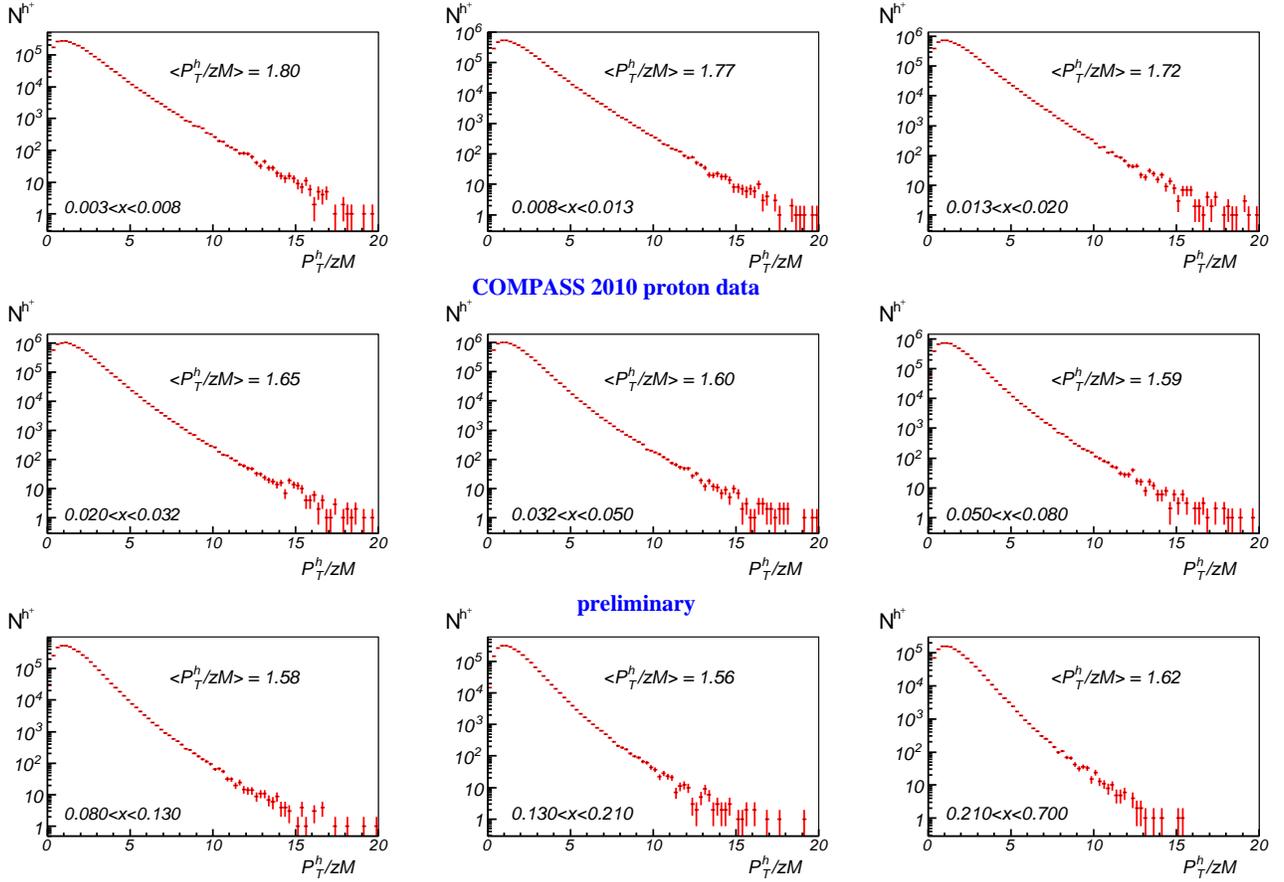}
\caption{Distributions of the weight factor $P_T/zM$ in the nine $x$ bins.}
% \end{minipage}
\label{fig:weights}
\end{figure*}

The distributions of the weight factor $P_T/zM$ in the
different $x$ bins are shown in 
 Fig. \ref{fig:weights} for the positive hadrons. 
The corresponding distributions for the negative hadrons are very similar. 
The $P_T/z$ acceptance of the spectrometer is about 60\% and rather flat.  
\begin{figure*}[tb]
\centering
\includegraphics[width=0.49\textwidth]{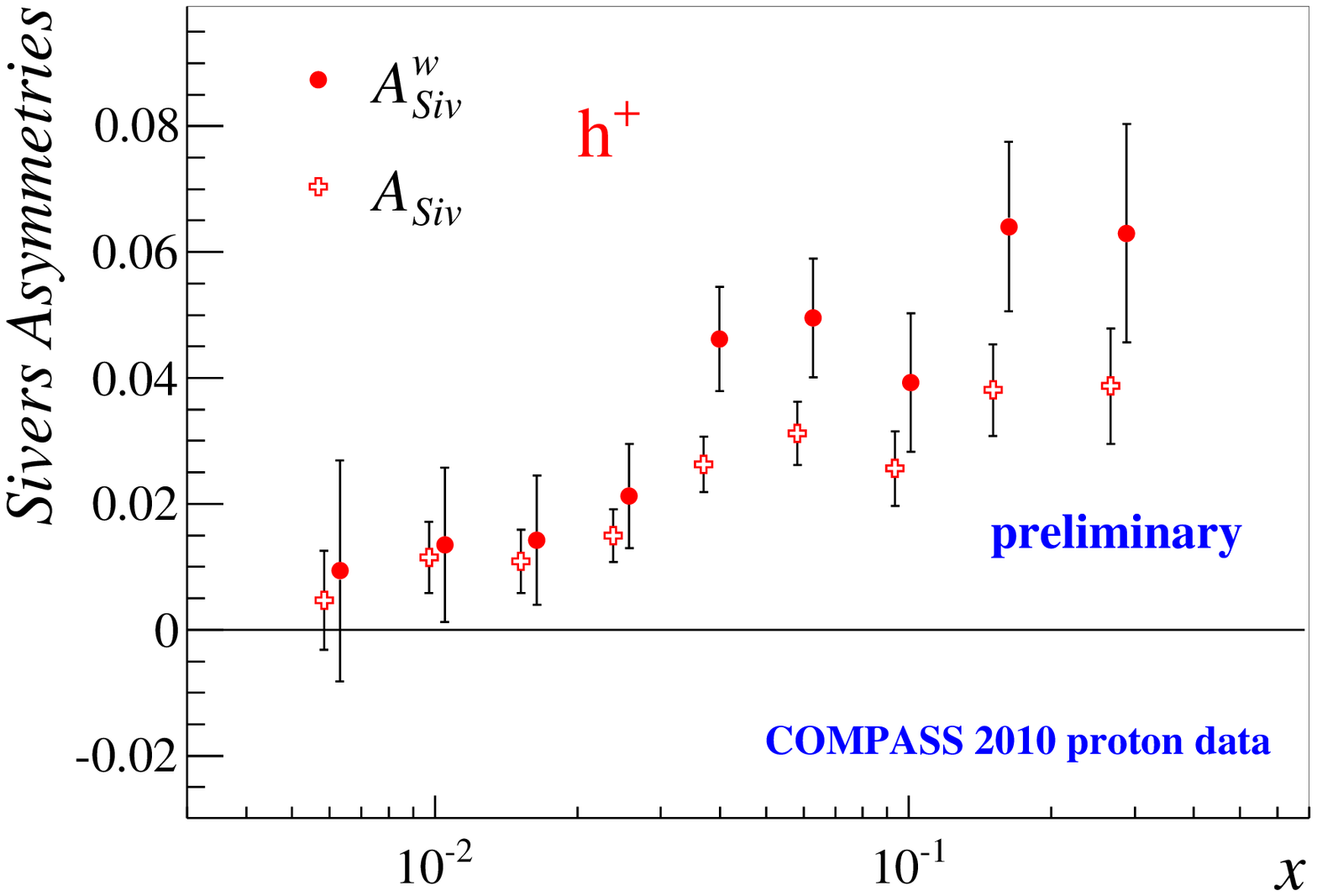}
\includegraphics[width=0.49\textwidth]{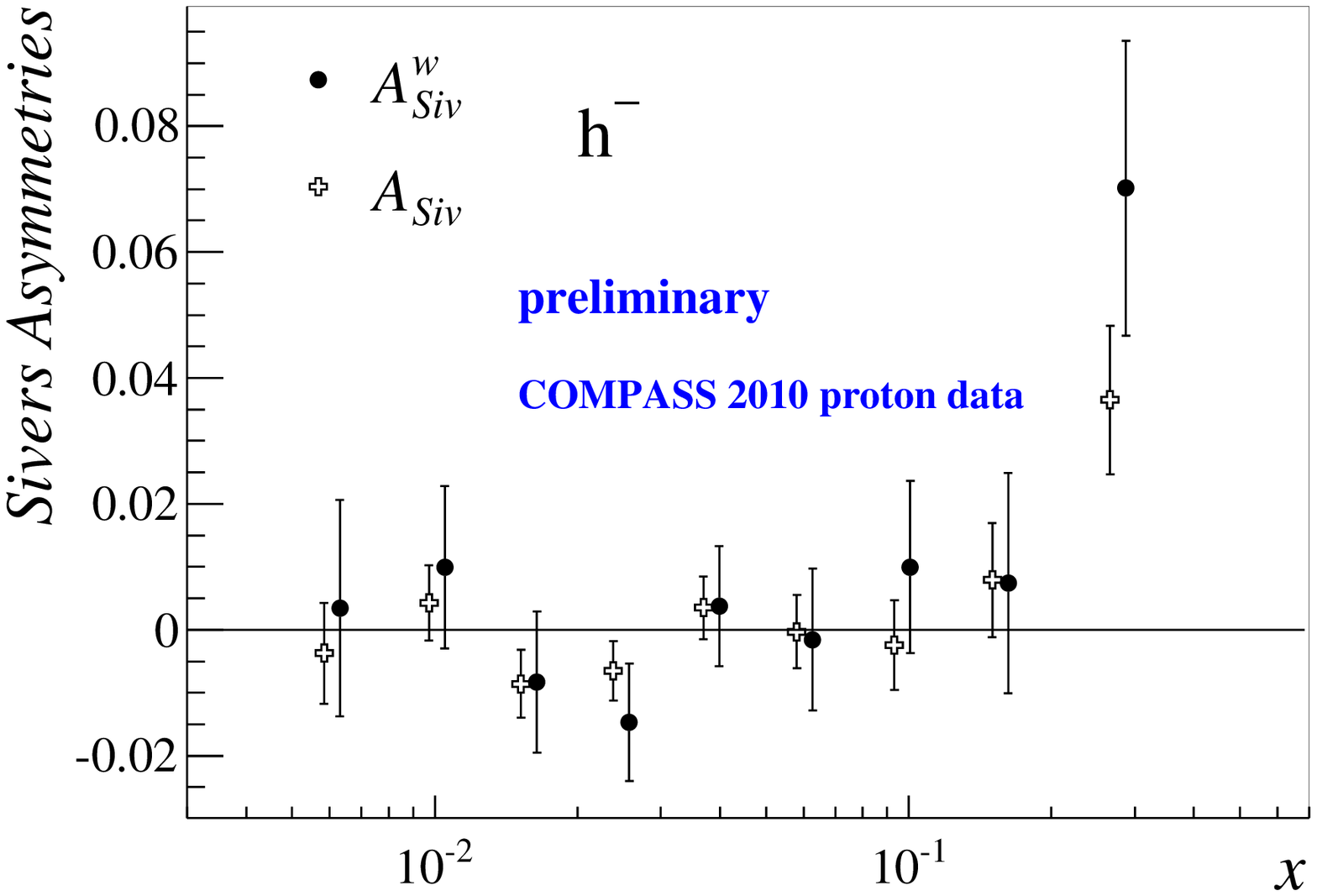}
\hfill
\caption{Full points: $A_{Siv}^{w}$ in the nine $x$ bins for positive 
(left panel) and negative (right panel) hadrons. The open crosses are the
standard Sivers asymmetries  $A_{Siv}$ from Ref. \cite{Adolph:2012sp}.}
\label{fig:results}
\end{figure*}

To extract $A_{Siv}^{w}$ we used an ad-hoc double-ratio method which 
utilizes the information coming from the different
cells  in which our target system 
is divided and insures cancellation of the target acceptance and of the beam 
flux. 
The method used in so far had to be modified since only the counts 
in the numerator of the expression of 
$A_{Siv}^{w}$  are weighted, while the 
counts at the denominator are unweighted. 
Three different estimators have been used, 
and it has been checked that the results are essentially 
identical. \\

The results in the case of positive hadrons are given in
Fig. \ref{fig:results}  left, while Fig. \ref{fig:results}
right gives the results for negative hadrons. 
In both cases the “standard” Sivers asymmetries published in 
\cite{Adolph:2012sp} are also 
plotted for comparison. 
As expected, the trend of the asymmetries is similar
both for positive and negative hadrons.
Assuming u-dominance, the results for positive hadrons 
which are clearly different from zero in particular at large $x$,
where $\langle Q^2 \rangle$ reaches $\sim 20$ GeV$^2$,
constitute the first direct measurement of 
$f_{1T}^{\perp \, (1) \, u}(x) / f_{1}^u(x)$.

\begin{figure}[tb]
\centering
\includegraphics[width=0.49\textwidth]{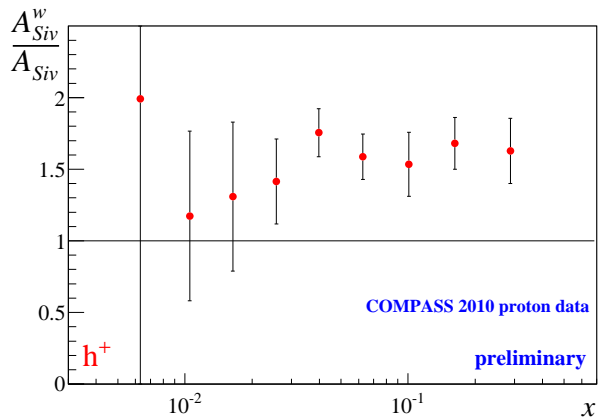}
\caption{The ratios $R^w$ between $A_{Siv}^{w}$ and $A_{Siv}^{w}$ in the nine
$x$-bins for positive hadrons.}
\label{fig:ratio_pos}
\end{figure}

Given the similar trend of the weighted asymmetries $A_{Siv}^{w}$ 
and of the standard asymmetries $A_{Siv}$, we have evaluated in each $x$ bin 
their ratios $R^w=A_{Siv}^{w}/A_{Siv}$. 
The values of $R^w$ 
are given in Fig. \ref{fig:ratio_pos} for positive hadrons, which exhibit 
a large Sivers asymmetry. 
In spite of the large statistical uncertainties
the ratios are compatible with a constant value, which is 
rather well determined ($1.6 \pm 0.1$). 
Also, the values in the different $x$ bins agree rather well with those of 
$\langle P_T / z M \rangle$, as expected.

To gain insight into the physics of the ratio $R^w$
it is useful to take for $A_{Siv}$ the expression  given in eq. (\ref{eq:gsa})
obtained in the
Gaussian model which from the past phenomenological analyses of
the Sivers effect is known to work rather well.
By comparison of eq. (\ref{eq:gsa}) and eq. (\ref{eq:wsa})
it is clear that their ratio provides some information on the quantity 
$a_G$.
In particular
 \begin{eqnarray}
R^w_G(x)&=&\frac{A_{Siv}^{w}(x)}{A_{Siv,G}(x)} \\
&=&\frac{2 }{\sqrt{\pi} M}
\frac{\sum_q e_q^2 x f_{1T}^{\perp \, (1) \, q}(x) \cdot \int dz D_1^q(z)}
{\sum_q e_q^2 x f_{1T}^{\perp \, (1) \, q}(x) \cdot \int dz z D_1^q(z) / \sqrt{ \langle P_T^2 \rangle _S} } \nonumber
\end{eqnarray}
can be further simplified by making the (reasonable) assumption
that
$\sqrt{\langle P_T^2 \rangle _S}
=\sqrt{z^2 \langle k_T^2 \rangle _S + \langle p^2_T \rangle}$
can be regarded as a constant since the dependence on $z$ of 
$\sqrt{\langle P_T^2 \rangle}$
is known to be much weaker than that of $D_1(z)$.
Further simplifications on the ratios $R^w_G$ can be obtained by weighting
the events by $P_T/M$ rather than $P_T/zM$.\\

As a conclusion, COMPASS is extracting the weighted Sivers asymmetry in 
the SIDIS process
of 160 GeV muons on transversely polarized protons
extending the standard measurements already performed.
Preliminary results from the data collected in 2010 have
 already been derived, using as weight $P_T/zM$ and are presented in 
this paper as a function of $x$.
No major experimental problems have been encountered and
the results look very promising in view of more precise extractions of
the Sivers function, and further measurements are foreseen.


\begin{thebibliography}{99} 

%\cite{Kotzinian:1994dv}
\bibitem{Kotzinian:1994dv}
  A.~Kotzinian,
  %``New quark distributions and semiinclusive electroproduction on the polarized nucleons,''
  Nucl.\ Phys.\ B {\bf 441} (1995) 234.

%\cite{Mulders:1995dh}
\bibitem{Mulders:1995dh}
  P.~J.~Mulders and R.~D.~Tangerman,
  %``The Complete tree level result up to order 1/Q for polarized deep inelastic leptoproduction,''
  Nucl.\ Phys.\ B {\bf 461} (1996) 197
   Erratum: [Nucl.\ Phys.\ B {\bf 484} (1997) 538].

%\cite{Bacchetta:2006tn}
\bibitem{Bacchetta:2006tn}
  A.~Bacchetta {\it et al.}, 
  %``Semi-inclusive deep inelastic scattering at small transverse momentum,''
  JHEP {\bf 0702} (2007) 093.

%\cite{Sivers:1989cc}
\bibitem{Sivers:1989cc}
  D.~W.~Sivers,
  %``Single Spin Production Asymmetries from the Hard Scattering of Point-Like Constituents,''
  Phys.\ Rev.\ D {\bf 41} (1990) 83.

%\cite{Airapetian:2004tw}
\bibitem{Airapetian:2004tw}
  A.~Airapetian {\it et al.} [HERMES Collaboration],
  %``Single-spin asymmetries in semi-inclusive deep-inelastic scattering on a transversely polarized hydrogen target,''
  Phys.\ Rev.\ Lett.\  {\bf 94} (2005) 012002.

%\cite{Airapetian:2009ae}
\bibitem{Airapetian:2009ae}
  A.~Airapetian {\it et al.} [HERMES Collaboration],
  %``Observation of the Naive-T-odd Sivers Effect in Deep-Inelastic Scattering,''
  Phys.\ Rev.\ Lett.\  {\bf 103} (2009) 152002.

%\cite{Alekseev:2010rw}
\bibitem{Alekseev:2010rw}
  M.~G.~Alekseev {\it et al.} [COMPASS Collaboration],
  %``Measurement of the Collins and Sivers asymmetries on transversely polarised protons,''
  Phys.\ Lett.\ B {\bf 692} (2010) 240.

%\cite{Adolph:2012sp}
\bibitem{Adolph:2012sp}
  C.~Adolph {\it et al.} [COMPASS Collaboration],
  %``II – Experimental investigation of transverse spin asymmetries in μ -p SIDIS processes: Sivers asymmetries,''
  Phys.\ Lett.\ B {\bf 717} (2012) 383.

%\cite{Adolph:2014zba}
\bibitem{Adolph:2014zba}
  C.~Adolph {\it et al.} [COMPASS Collaboration],
  %``Collins and Sivers asymmetries in muonproduction of pions and kaons off transversely polarised protons,''
  Phys.\ Lett.\ B {\bf 744} (2015) 250.

%\cite{Qian:2011py}
\bibitem{Qian:2011py}
  X.~Qian {\it et al.} [Jefferson Lab Hall A Collaboration],
  %``Single Spin Asymmetries in Charged Pion Production from Semi-Inclusive Deep Inelastic Scattering on a Transversely Polarized $^3$He Target,''
  Phys.\ Rev.\ Lett.\  {\bf 107} (2011) 072003.

%\cite{Alexakhin:2005iw}
\bibitem{Alexakhin:2005iw}
  V.~Y.~Alexakhin {\it et al.} [COMPASS Collaboration],
  %``First measurement of the transverse spin asymmetries of the deuteron in semi-inclusive deep inelastic scattering,''
  Phys.\ Rev.\ Lett.\  {\bf 94} (2005) 202002.

%\cite{Ageev:2006da}
\bibitem{Ageev:2006da}
  E.~S.~Ageev {\it et al.} [COMPASS Collaboration],
  %``A New measurement of the Collins and Sivers asymmetries on a transversely polarised deuteron target,''
  Nucl.\ Phys.\ B {\bf 765} (2007) 31.

%\cite{Alekseev:2008aa}
\bibitem{Alekseev:2008aa}
  M.~Alekseev {\it et al.} [COMPASS Collaboration],
  %``Collins and Sivers asymmetries for pions and kaons in muon-deuteron DIS,''
  Phys.\ Lett.\ B {\bf 673} (2009) 127.

%\cite{Efremov:2003tf}
\bibitem{Efremov:2003tf}
  A.~V.~Efremov, K.~Goeke and P.~Schweitzer,
  %``Sivers versus Collins effect in azimuthal single spin asymmetries in pion production in SIDIS,''
  Phys.\ Lett.\ B {\bf 568} (2003) 63.

%\cite{Vogelsang:2005cs}
\bibitem{Vogelsang:2005cs}
  W.~Vogelsang and F.~Yuan,
  %``Single-transverse spin asymmetries: From DIS to hadronic collisions,''
  Phys.\ Rev.\ D {\bf 72} (2005) 054028.

%\cite{Collins:2005ie}
\bibitem{Collins:2005ie}
  J.~C.~Collins {\it et al.}, 
  %``Sivers effect in semi-inclusive deeply inelastic scattering,''
  Phys.\ Rev.\ D {\bf 73} (2006) 014021.

%\cite{Anselmino:2012aa}
\bibitem{Anselmino:2012aa}
  M.~Anselmino, M.~Boglione and S.~Melis,
  %``A Strategy towards the extraction of the Sivers function with TMD evolution,''
  Phys.\ Rev.\ D {\bf 86} (2012) 014028.


%\cite{Kotzinian:1995cz}
\bibitem{Kotzinian:1995cz}
  A.~M.~Kotzinian and P.~J.~Mulders,
  %``Longitudinal quark polarization in transversely polarized nucleons,''
  Phys.\ Rev.\ D {\bf 54} (1996) 1229.

%\cite{Kotzinian:1997wt}
\bibitem{Kotzinian:1997wt}
  A.~M.~Kotzinian and P.~J.~Mulders,
  %``Probing transverse quark polarization via azimuthal asymmetries in leptoproduction,''
  Phys.\ Lett.\ B {\bf 406} (1997) 373.

%\cite{Boer:1997nt}
\bibitem{Boer:1997nt}
  D.~Boer and P.~J.~Mulders,
  %``Time reversal odd distribution functions in leptoproduction,''
  Phys.\ Rev.\ D {\bf 57} (1998) 5780.


%\cite{Gregor:2005qv}
\bibitem{Gregor:2005qv}
  I.~M.~Gregor [HERMES Collaboration],
  %``Transverse spin physics at HERMES,''
  Acta Phys.\ Polon.\ B {\bf 36} (2005) 209.

%\cite{Kang:2012ns}
\bibitem{Kang:2012ns}
  Z.~B.~Kang, I.~Vitev and H.~Xing,
  %``Transverse momentum-weighted Sivers asymmetry in semi-inclusive deep inelastic scattering at next-to-leading order,''
  Phys.\ Rev.\ D {\bf 87} (2013) no.3,  034024.

\bibitem{Sbrizzai2016}
G. Sbrizzai
for the COMPASS Collaboration, ``TMD measurements at COMPASS'' 
QCD-N – 4rd Workshop on the QCD Structure of the Nucleon, Getxo, Bilbao, Spain, 11-15 July, 2016.

\bibitem{baronespin}
A. Martin, F. Bradamante and V. Barone, these proceedings, and 
  arXiv:1701.08283 [hep-ph].

\end{thebibliography}
\end{document}